\documentclass[aps,pre,twocolumn,showpacs,superscriptaddress,groupedaddress]{revtex4-1}  % for 
\usepackage{graphicx}  % needed for figures
\usepackage{dcolumn}   % needed for some tables
\usepackage{bm}        % for math
\usepackage{amsmath,amssymb}   % for math

\begin{document}

\title{{Transition to a labyrinthine phase in a driven granular medium}}
\affiliation{Universit\'e Paris Diderot, Sorbonne Paris Cit\'e, MSC, CNRS (UMR 7057), 75013 Paris, France}
\author{Simon Merminod} \affiliation{Universit\'e Paris Diderot, Sorbonne Paris Cit\'e, MSC, CNRS (UMR 7057), 75013 Paris, France}
\author{Timoth\'ee Jamin} \affiliation{Universit\'e Paris Diderot, Sorbonne Paris Cit\'e, MSC, CNRS (UMR 7057), 75013 Paris, France}
\author{Eric Falcon} \affiliation{Universit\'e Paris Diderot, Sorbonne Paris Cit\'e, MSC, CNRS (UMR 7057), 75013 Paris, France}
\author{Michael Berhanu} \affiliation{Universit\'e Paris Diderot, Sorbonne Paris Cit\'e, MSC, CNRS (UMR 7057), 75013 Paris, France}

\begin{abstract} 
Labyrinthine patterns arise in two-dimensional physical systems submitted to competing interactions, ranging from the fields of solid-state physics to hydrodynamics. For systems of interacting particles, labyrinthine and stripe phases were studied in the context of colloidal particles confined into a monolayer, both numerically by means of Monte Carlo simulations and experimentally using superparamagnetic particles. Here we report an experimental observation of a labyrinthine phase in an out-of-equilibrium system constituted of macroscopic particles. Once sufficiently magnetized, they organize into short chains of particles in contact and randomly orientated. We characterize the transition from a granular gas state towards a solid labyrinthine phase, as a function of the ratio of the interaction strength to the kinetic agitation. Spatial local structure is analyzed by means of an accurate particle tracking. Moreover, we explain the formation of these chains using a simple model. 
\end{abstract}

\maketitle
\section{Introduction}
Labyrinthine phases are intriguing two-dimensional patterns occurring in various domains of physics, in equilibrium and out-of-equilibrium situations. Two distinct phases form at small-scale well separated stripes, which are themselves entangled, leading to a complex large-scale pattern. These shapes were experimentally obtained for extremely varied two-dimensional (2D) systems ranging from ferrimagnetic garnet films~\cite{SeulScience91} in condensed matter and Langmuir monolayer~\cite{SeulPRL90} in soft matter, to granular fluid suspension in which air penetrates~\cite{SandnesPRL07,SandnesNatcom2011}, ferrofluid drop~\cite{DicksteinScience1993} and biphasic ferrofluid-oil layer~\cite{EliasJPF97} in fluids mechanics, and chemical reaction-diffusion systems~\cite{LeePRE1995}. The common denominator of these systems is the competition between long-range repulsion and short-range attraction, which leads to the phenomenology of modulated phases~\cite{SeulScience95}. Moreover, a wide range of ordering effects that lead to different patterns can also be related to the competition between interactions and geometrical frustration, as specifically shown for magnetic thin films~\cite{De'Bell2000}.

\indent By analogy with the phenomenology of these continuous systems, labyrinthine and stripe phases have been introduced for systems of particles. In particular in the context of colloidal monolayers, several Monte-Carlo simulations~\cite{MalescioNatMat03,CampPRE03,Dobnikar2008,ShokefPRL09} and one molecular dynamics simulation~\cite{HawPRE2010} have been performed. 
It was shown that, to observe stripes and labyrinthine phases, a long-range repulsive potential is needed, together with a short-range attraction which can be replaced by a core-softened potential~\cite{MalescioNatMat03,CampPRE03}. Tuning geometrical frustration in non-interacting colloidal monolayers~\cite{ShokefPRL09,HanNature08} leads also to stripe phases. Moreover the only experimental observation of a labyrinthine phase in a colloidal system was obtained using superparamagnetic colloids under a magnetic field, inducing dipolar interactions~\cite{OstermanPRL07}. Labyrinthine phases were indeed found as equilibrium states at high enough density of micrometric spheres in agreement with dedicated Monte-Carlo simulations~\cite{Dobnikar2008,OstermanPRL07}. In contrast, similar labyrinthine or stripe phases have not been described in a macroscopic and out-of-equilibrium system whose particles can be individually identified.
\begin{figure}[!t]
\centering
\includegraphics[width=.39\textwidth]{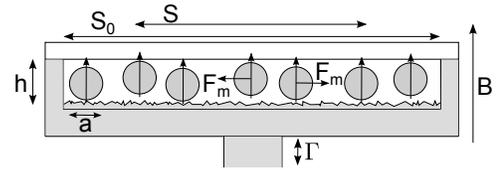}
\caption{Experimental setup. The $5000$ chromed steel spherical particles (diameter~$a=1$\,mm and mass $m=4.07$\;mg) are vertically vibrated (acceleration~$\Gamma = 21.9\,$m.s$^{-2}$) inside a horizontal, square aluminium cell (area $S_0=9\times 9$\,cm$^2$) with a rough bottom plate and a polycarbonate top lid (gap size $h=1.5\,a$). In presence of a vertical magnetic field~$B$, particles repel each other with a force $\vec{F_m}$. The region of interest is of area $S=5.7\times 5.7$\,cm$^2$.}
\label{fig:f1}
\end{figure}

\begin{figure*}[!t]
\centering
\includegraphics[width=1\textwidth]{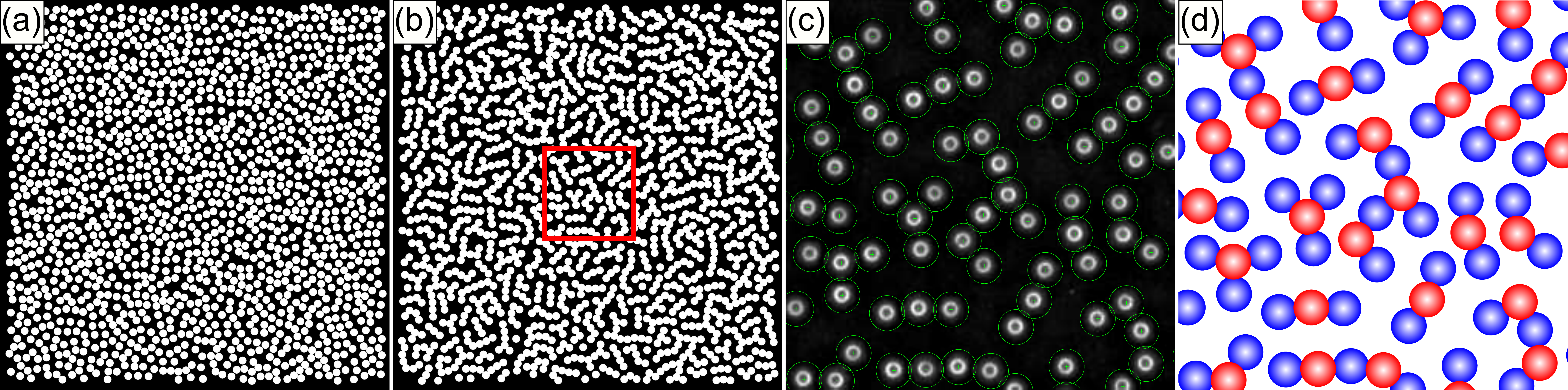}
\caption{(color online) Top views of the system of particles. In (a) and (b), images of the spheres have been replaced by white disks of diameter $a$ for better visualization, whereas (c) is from a direct image from the camera and (d) is the result of particle tracking. (a)~Dissipative granular gas state at moderate $B$ ($80\,$G). (b)~Labyrinthine phase at high $B$ ($170\,$G). The particles organize into an amorphous phase mostly composed of chains of a few particles. The region within the red square is enlarged in (c) and (d). (c)~Thick circles are reflections of lighting on the spheres and appear smaller than the actual overlapping particle radii (thin circles). (d)~Buckled chains (red or light gray for top particles, blue or dark gray for bottom particles) mostly compose the amorphous phase.}
\label{fig:f2}
\end{figure*}

\indent Here, we report the observation of such a labyrinthine phase. A monolayer of soft ferromagnetic spheres of 1 mm in diameter is vibrated to form a 2D granular gas~\cite{Olafsen1998,Olafsen1999,Losert1999,Moon2001,ReisPRE2007,JaegerRMP96}. Under mechanical agitation particles undergo a Brownian-like motion, but due to the dissipative nature of the collisions, the granular gas reaches a stationary out-of-equilibrium state. Then immersed in a vertical external magnetic field $B$, the soft ferromagnetic spheres are magnetized and interact with each other as induced dipoles. When the magnetic field is increased, the granular gas solidifies into a phase composed of chains of a few particles in contact, similar to the labyrinthine phase observed with colloids~\cite{OstermanPRL07}. In contrast to this colloid study which focuses on equilibrium states, the transition from gas to labyrinth is here clearly described, using an accurate particle tracking. Finally as a remark, we emphasize that the physical mechanisms at play in labyrinthine and stripe phases of interacting particles, differ from those in chain and cluster phases reported in some interacting granular gases~\cite{Blair2003,SnezhkoPRL2005,Oyarte2013}, despite visual similarities. Indeed, these phases are composed of head-to-tail dipoles and were observed when attractive behavior is dominant at large scale, because permanent dipoles are considered~\cite{Blair2003,Oyarte2013} or because the hypothesis of a quasi-2D system is not verified~\cite{SnezhkoPRL2005}.

\section{From a granular gas towards a labyrinthine solid phase}

\indent The experimental setup is similar to the one used in~\cite{MerminodEPL2014}. Soft ferromagnetic spherical particles of diameter $a=1\,$mm are confined between two horizontal parallel plates separated by a gap $h=1.5\,a$ in order to form a monolayer as depicted in Fig.~\ref{fig:f1}. Particles are vibrated vertically and are lightened by an annular LED array and imaged from the top by a fast camera through the transparent top plate. Particle center positions are tracked and their trajectories are reconstructed in the horizontal plane~\cite{ReisPRE2007,Crocker}. Interactions between particles are introduced by means of an external vertical magnetic field of amplitude $B$ controlled by the experimentalist. Additional details on the experimental setup and protocol, and on the particle detection technique, are given in the Appendices.

\indent Magnetized soft ferromagnetic spheres behave as induced dipoles, whose magnetic moments are vertical and proportional to $B$. The interaction potential $U_m$ of a particle located at a distance $r$ and a polar angle $\theta$ from a second particle~\cite{Jackson} reads in spherical coordinates~\footnote{
See Appendix A for complements on the formula of $U_m$.}, with $\mu_0$  the vacuum permeability:
\begin{equation}
U_m  (r,\theta) = -\frac{\pi}{16} \frac{{B}^2}{\mu_0} \frac{a ^6}{r^3} \left( 2\cos ^2\theta -\sin ^2\theta \right)
\label{eq:Em}
\end{equation}

Two spheres in the same horizontal ($\theta=\pi/2$) plane are thus repelling each other. Using this experimental method, macroscopic transitions were observed in 3D assemblies of magnetized soft-ferromagnetic particles~\cite{LarocheEPJB10,LopezPRL10}. Then, the number of particles per surface unit is expressed by a dimensionless parameter, the area fraction $\phi=({N \pi a^2})/({4\,S)}$, with $N$ the number of particles tracked in the region of interest ${S}$. For a monolayer of particles, high enough magnetic field and moderate area fraction ($\phi=0.2$), the 2D granular gas solidifies into a hexagonal crystal~\cite{MerminodEPL2014}, whose melting has been found to follow the Kosterlitz-Thouless-Halperin-Nelson-Young (KTHNY) scenario~\cite{SchockmelPRE2013}.\\
\indent Here $\phi$ is increased to $0.5$. For moderate values of the magnetic field $B$ and continuous shaking, particles undergo a Brownian-like motion. At a given instant particle positions are random [Fig.~\ref{fig:f2}(a)] and spheres exchange energy through dissipative collisions and magnetic interactions. We observe a 2D granular gas, whose properties are similar to those found at lower area fraction~\cite{MerminodEPL2014}. Then by increasing $B$ further, we observe that despite the magnetic repulsion, small chains of two or three particles in contact are starting to form in the bulk of agitated particles. We remark also that the motions of the particles belonging to these chains are considerably restricted compared to those of the free particles. At higher magnetic field, the quasi-totality of the particles are condensed into these chains [Fig.~\ref{fig:f2}(b)]. At large-scale, the picture of the assembly of the system does not present an ordered structure. Thus, by increasing magnetic interactions, the system has been solidified in an amorphous state. Labyrinthine patterns were indeed described as globally disordered stripe domains~\cite{SeulPhilMagB1992}. Due to the presence of chains, the particle assembly presents striking similarities with the labyrinthine and stripe phases observed~\cite{OstermanPRL07} and numerically predicted~\cite{MalescioNatMat03,CampPRE03,Dobnikar2008,ShokefPRL09,HawPRE2010} for example for 2D colloidal systems under thermal agitation and with dipolar repulsive magnetic interactions. Here, a transition from a granular gas phase to a labyrinthine phase for a macroscopic, out-of-equilibrium system is observed. This transition can be also visualized by applying a linearly increasing magnetic field from $B=0\,$G to $B=200\,$G (see Movie in Supplemental Material~\footnote{see Supplemental Material at [URL will be inserted by publisher] for a movie of the transition}). First, pairs aggregate, then triplets, and so on homogeneously across the cell, until nearly motionless chains of various lengths occupy the whole cell, isolating the few remaining fluctuating particles from each other. We notice also that, starting from the labyrinthine phase, the inverse transition is observed when the magnetic field is decreased. This shows that the system does not present any noticeable hysteresis.\\
\indent By the means of an accurate particle tracking, chain morphology can be now quantitatively characterized. Chains are well separated due to the magnetic repulsion and can be thus considered as groups of more than one particle, according to a criterion of the contact distance. Moreover chains appear mainly as linear objects because, most of the time, a particle inside a chain is in contact with two neighboring particles. Nevertheless the relative orientations of the chains seem random. In the following, quantitative analysis of the small-scale structure will reveal that chains correspond to a buckled state of particles in contact [Fig.~\ref{fig:f2}(c)]. A particle, once it is condensed in a chain, is either in contact with the top plate or with the bottom plate. Using slight differences of lighting for the two kinds of particles, our detection technique is able to provide the vertical position of the particles in the chains [as shown in virtual image Fig.~\ref{fig:f2}(d)], that is coded as \emph{up} (red) or \emph{down} (blue). It can be noticed that the particles at the tips of the chains are for the most part \emph{down}, revealing an effect of gravity.

\section{Characterization of the transition}

\begin{figure}[!t]
\centering
\includegraphics[width=.34\textwidth]{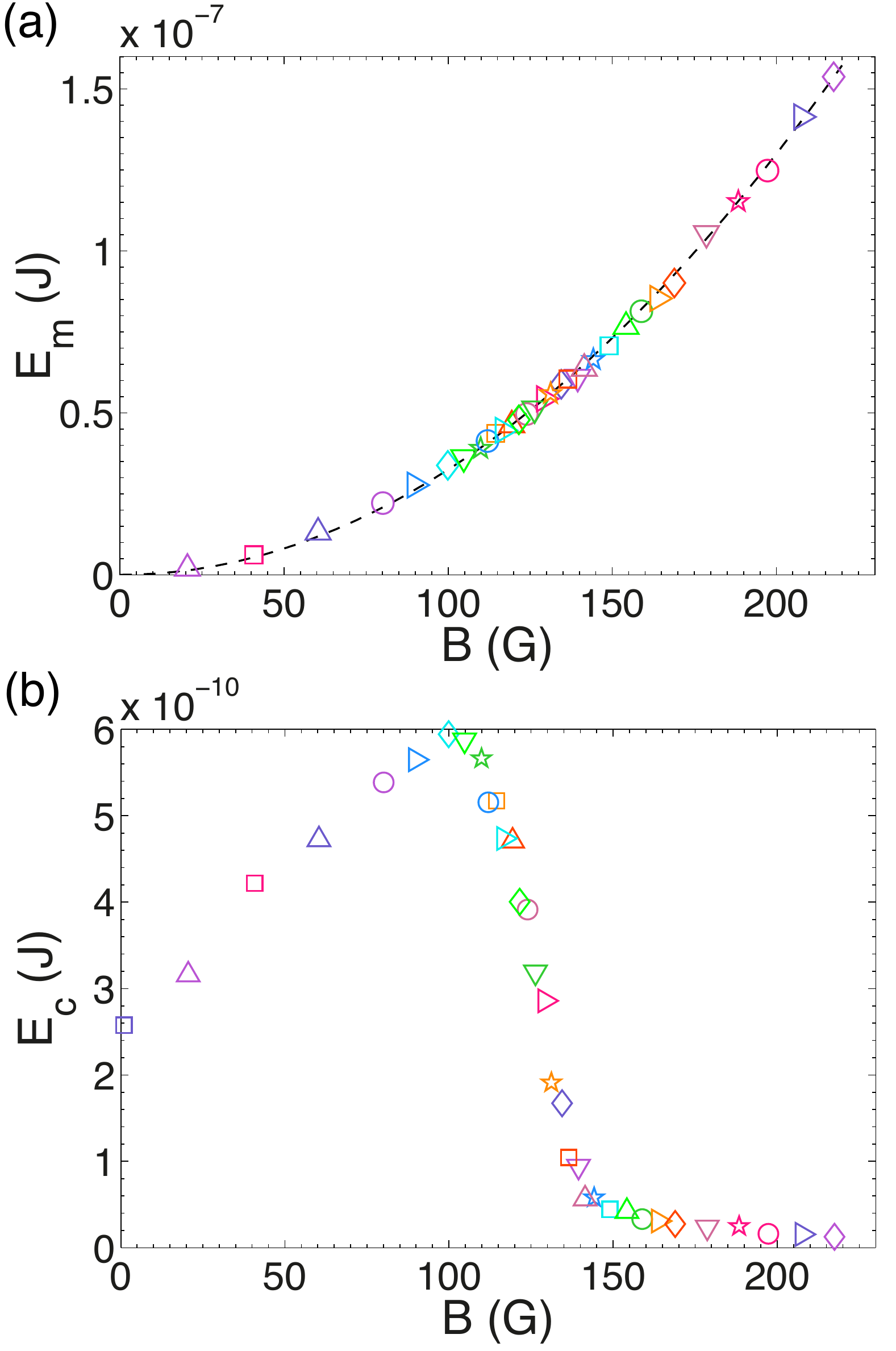}
\caption{(color online) Potential magnetic energy and kinetic energy per particle. Every marker corresponds to an independent experiment. (a)~The potential magnetic energy is found to scale as $B^2$ (the dashed curve is a $B^2$ fit). (b)~The kinetic energy, which measures agitation, plotted as a function of $B$. First, $E_c$ increases due to the fluidizing effect of the magnetic interactions \cite{MerminodEPL2014}, then it suddenly drops towards zero at the onset of the solidification ($B\approx 100$\,G).
}
\label{fig:f2s}
\end{figure}

Let us now quantitatively characterize the transition from granular gas to the labyrinthine phase using the particle tracking data. The magnetic potential energy $E_m$ and the horizontal kinetic energy $E_c$ per particle can be now computed. From the interaction potential $U_m(r,\theta)$ defined in Eq.~(\ref{eq:Em}), the magnetic energy per particle is computed as the averaged summation over the pairs of the interaction potential of each pair: 
\begin{equation}
E_{m} = \overline{\frac{1}{N_p} \, \sum_{i=1}^{N_p}\sum_{j=i+1}^{N_p} U_m(r_{ij},\theta_{ij})}
\label{eq:Em2}
\end{equation}
with $N_p$ the number of particles involved in the calculation of $E_m$, $r_{ij}$ and $\theta_{ij}$ respectively the distance and the polar angle between the two particles $i$ and $j$, and $\overline{\;\cdot\;\rule[0mm]{0mm}{2mm}}$ the temporal average. The magnetic potential energy depends on the local configuration of the particles, and therefore it fluctuates in time. Its averaged value, $E_m$, is found to be proportional to $B^2$ [Fig.~\ref{fig:f2s}(a)]. The kinetic energy per particle is computed from velocity measurements:
\begin{equation}
E_c=\frac{1}{2}\,m\, \overline{( \langle {v_x}^2\rangle +\langle {v_y}^2\rangle \rule[0mm]{0mm}{4mm})}
\label{eq:Ec}
\end{equation}
where $m$ denotes the particle mass, $v_x$ (resp. $v_y$) the horizontal velocities in the $x$-direction ($y$-direction) and $\langle\cdot\rangle$ an ensemble average. $E_c$ is a measure of the agitation in the system.
When the magnetic field $B$ is increased, in a first stage $E_c$ grows [Fig.~\ref{fig:f2s}(b)]. Repulsive dipole-dipole interactions reduce the rate of dissipative collisions, which increases consequently $E_c$ for a given shaking amplitude~\cite{MerminodEPL2014}. Once chains start to form, for $B\approx 100\,$G, $E_c$ drops significantly and nearly vanishes as the labyrinthine phase is formed for $B\approx 150\,$G, illustrating the solidification process. For higher magnetic excitation values, the labyrinthine phase becomes less and less mobile as interactions strengthen and fluctuations are restrained.

\begin{figure*}[!t]
\centering
\includegraphics[width=1\textwidth]{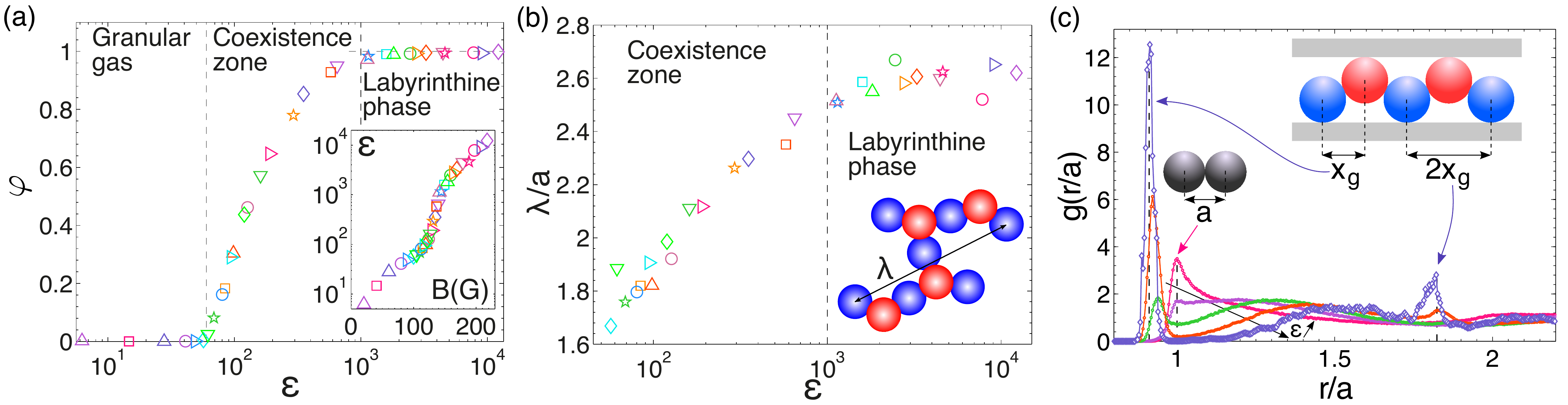}
\caption{(color online) Characterizing the transition. Every marker corresponds to an independent experiment.
 (a)~The main plot shows the fraction of particles in chains~$\varphi$, as a function of the control parameter $\varepsilon = E_m /E_c$. It varies from about zero in the dissipative state to one in the fully solidified state. The horizontal, black dashed line is placed as a guide. The inset shows $\varepsilon$ as a function of $B$. (b)~Adimensionalized mean chain length $\lambda/a$ as a function of $\varepsilon$, with $\lambda$ evaluated by computing the largest distance between particle centers inside a given chain. The value of $\lambda$ averaged over the chains increases continuously with $\varepsilon$. If the averaging is weighted by the number of particles in the chains, the obtained values are larger by roughly one diameter unit, but the trend with $\varepsilon$ is similar. (c)~Evolution of the radial pair correlation function $g(r/a)$ for growing values of $\varepsilon$ (see arrow): $\varepsilon=14.7$ (red), $\varepsilon=41.0$ (purple), $\varepsilon=68.9$ (green), $\varepsilon=127$ (orange) and $\varepsilon=649$ (blue). As the transition occurs, the $r=a$ peak of $g(r/a)$ drops down to zero, showing that in-plane collisions probability vanishes for high values of $\varepsilon$. In the meantime, another sharp peak grows at $r/a=x_g/a=0.91$, which is the imprint of the buckled chains.}
\label{fig:f3}
\end{figure*}
\indent Now, let us define the dimensionless control parameter $\varepsilon \equiv E_m/E_c$~\cite{MerminodEPL2014}, which is depicted as a function of $B$ in the inset of Fig.~\ref{fig:f3}(a). $\varepsilon$ provides a measure of the competition between distance interactions and kinetic agitation. By analogy with an order parameter, the fraction of particles condensed in the chains $\varphi$ is computed as the ratio of the number of particles belonging to a group of more than one particle to the total number of particles tracked in the region of interest. By plotting $\varphi$ as a function of $\varepsilon$, as shown in Fig.~\ref{fig:f3}(a), the transition is well depicted. For $\varepsilon < 60$, $\varphi$ is nearly null in the granular gas phase, whereas for $ \varepsilon >1000$, $\varphi$ is slightly smaller than one, for the labyrinthine phase. The intermediate region of partial solidification corresponds to a coexistence zone between fluidized particles and particles condensed in the chains.
Let us emphasize that, although the simple criterion of the variation of $\varphi$ captures well the transition from a granular gas to a labyrinthine phase, it should fail to distinguish a stripe phase from a labyrinthine phase. Topology and morphology would indeed have to be taken into account, like local orientational properties. Several approaches were proposed to analyze or to model labyrinthine patterns, such as the introduction of a local wave vector~\cite{LeBerrePRE2002}, the computation of the wrinkledness~\cite{ReimannPRE2002}, and the decomposition of the pattern into clusters of linear segments~\cite{SeulPhilMagB1992}. To our knowledge, the definition of an appropriate order parameter for labyrinthine patterns remains an open question.

Nonetheless, aiming at quantifying some of the directly observable morphological changes of the chains, we evaluate their mean length $\lambda$ as a function of $\varepsilon$ [Fig.~\ref{fig:f3}(b)].
$\lambda$ is defined as the average over all chains of the largest distance between particle centers inside a given chain.
Starting from $1.6\,a$ at the formation of first chains, $\lambda$ seems to saturate for the highest values of $\varepsilon$ around $2.6\,a$. This suggests that competition between growing chains could limit their extension.\\

\indent The pair correlation function, which is related to the probability of finding a particle center at a given distance from another particle center, provides information on the small-scale structure of the system. Before the transition, the particle assembly evolves from a purely dissipative gas to an effectively more elastic gas~\cite{MerminodEPL2014}. Therefore, as Fig.~\ref{fig:f3}(c) (red and purple lines) displays, the peak at the diameter value flattens whilst the effective elasticity rises up. From the onset of solidification (green, orange and blue lines), surprisingly an extremely sharp peak grows from zero at the distance value $r=x_g\approx 0.91\,a$, which is smaller than the particle diameter. This peak, which would be impossible to observe in purely 2D systems of hard spheres, reveals the internal structure of the chains. Here the gap size is indeed large enough so that partial overlaps of particles are allowed, leading to the formation of buckled particle chains in which particles are in contact with the top or bottom plate [see schematic in Fig.~\ref{fig:f3}(c)]. Geometrical calculations yielding $x_g=\sqrt{2h\,a-h^2\,}$, one finds $h = 1.42\,$mm, which corresponds to the announced gap of $1.5\,$mm diminished by the the roughness of the bottom plate.
Moreover in the labyrinthine phase (blue line), $g(r)$ exhibits also a shorter peak at the position $2\,x_g$, from the aligned second neighbors, showing the presence of linear chains. Between these two peaks, a zero-probability at $ x_g \lesssim r \lesssim 1.2\,a$ indicates the void spaces between the chains while the non-vanishing probability for $ 1.2\,a \lesssim r \lesssim 2 x_g$ stands for both non-aligned second neighbors in chains and particles from neighboring chains. The 3D effects related to the gap size $h$ are thus essential to describe the small-scale structure of the labyrinthine phase. Therefore, we discuss now how three-dimensionality can explain the stability of the chains at high enough area fraction.

\begin{figure}
\centering
\includegraphics[width=.45\textwidth]{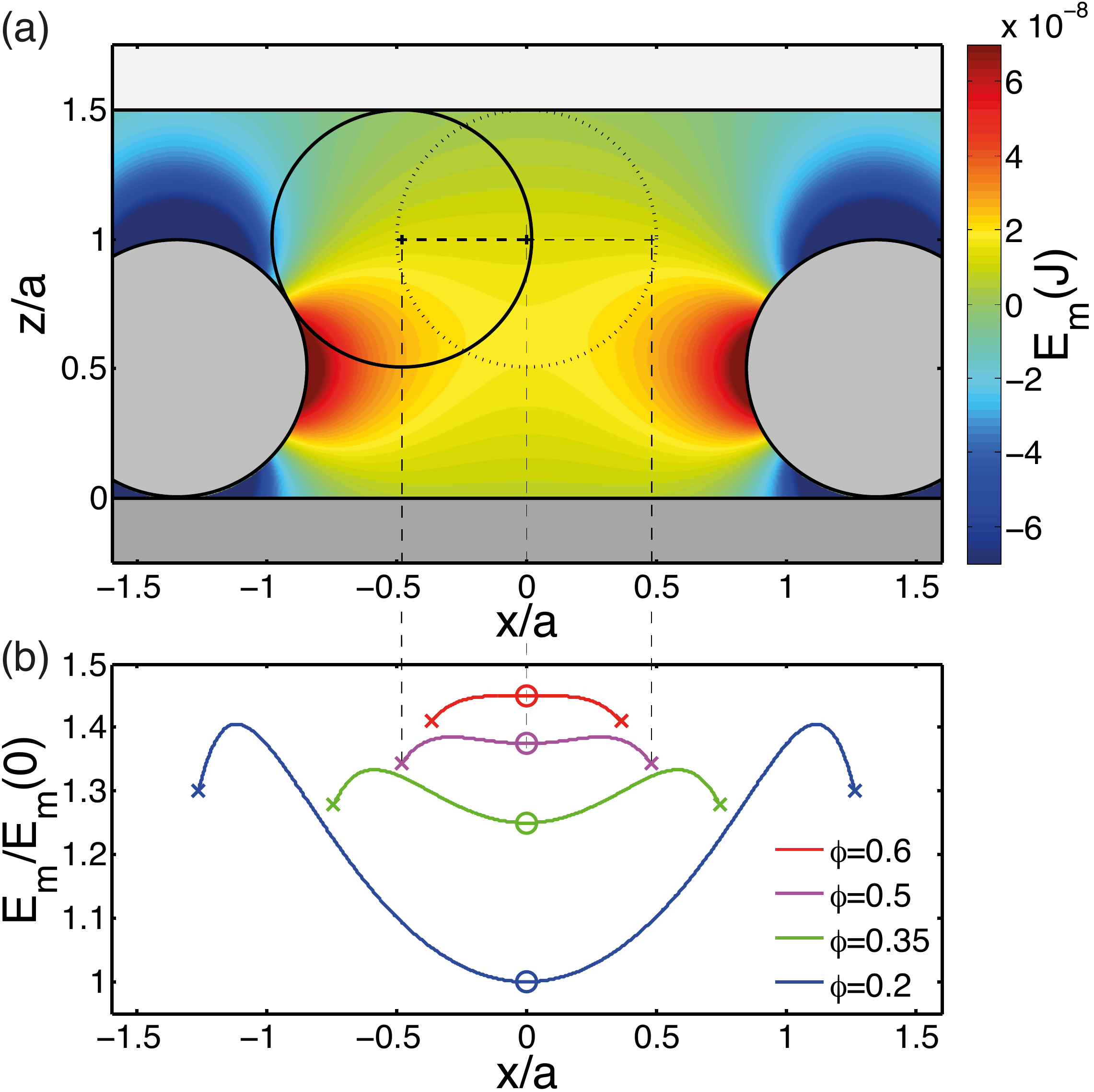}
\caption{(color online) (a)~The potential energy landscape $E_m$ is computed for $\phi=0.5$ and $B=200\,$G by averaging the pair potential $U_m$ [see Eq.~(\ref{eq:Em})] for six \emph{down} neighbors forming a hexagon (four particles are out of the figure plane) and a central \emph{up} particle moving along the $x$-axis (dashed line). (b)~The profile of $E_m$ is plotted along this trajectory for four values of $\phi$ (curves have been rescaled and shifted vertically for clarity). Circle mark depicts the initial central position and cross marks show the two contact positions. From these graphs, for $\phi=0.2$ and $0.35$ hexagonal configurations are found stables, whereas for $\phi=0.5$ and $0.6$, the central particle in presence of agitation should leave the position $x=0$ to reach contact positions associated with the buckled chains.
}
\label{fig:f4}
\end{figure}
\section{Chain formation mechanism}
\indent  At low area fraction ($\phi=0.2$ and lower), the stable state of the assembly of spheres in dipolar interaction was found to be a hexagonal crystal~\cite{MerminodEPL2014,SchockmelPRE2013}. Why does the hexagonal structure become now unstable at higher area fraction and how can we explain the formation of chains of particles in contact? In Fig.~\ref{fig:f4}(a), we plot the 3D magnetic energetic landscape (in a vertical plane) for a sphere initially in the center of a hexagon of 6 neighbouring particles, the projected horizontal distance between the particles being geometrically given by $d=a\,\sqrt{\pi/(2\sqrt{3}\phi)}$. Let us consider this central particle at $x=0$ assumed \emph{up} [dashed circle in Fig.~\ref{fig:f4}(a)] between six \emph{down} neighbors all at a distance $d$, and investigate its potential energy when it moves from $x=0$ to the contact position for several values of $\phi$ [Fig.~\ref{fig:f4}(b)]. Contact positions between spheres are local minima of potential energy, as the dipolar interaction aims to align spheres along a vertical axis. For $\phi=0.2$, the central position is an absolute minimum of energy, in agreement with the expected stability of the hexagonal lattice. In contrast when $\phi$ is increased, the energy barrier decreases and $E_m(0)$ augments relatively to $E_m(\pm d)$. For $\phi=0.5$ contact positions become absolute minima, which can be reached by means of mechanical agitation. For $\phi=0.6$, the central position is not even a minimum anymore. Therefore for $\phi=0.5$ and above, we expect that hexagonal structure is unstable, leading to local structures of spheres in contact like chains, despite the isotropic dipolar repulsion in a purely 2D system. This qualitative model explains the small-scale attraction leading to particle contacts, needed for the shaping and the stability of labyrinthine phases~\cite{SeulScience95}. To improve the description, solid friction between the spheres and the top and bottom plates should be incorporated, as it may greatly stabilize the buckled chains. In thermal systems, similar predictions were obtained using Monte-Carlo simulations~\cite{MalescioNatMat03,CampPRE03,Dobnikar2008}. In both these examples and in our system, the resultant of repulsive interactions of the assembly of particles over one acts as a magnetic pressure, favoring contact at high enough particle density.
Additionally, we notice that buckled phases stabilized by pressure and friction can also appear in thin vibrated granular layers without magnetic interactions~\cite{Melby2005} if the density and gap size are sufficiently large \footnote{In our cell with $h=1.5\,a$, the total number of particles should be larger than $9477$ for such a buckled state to be observed~\cite{Melby2005}.}. Nevertheless, in this last case, the structuring in separated chains is absent.

\section{Conclusion}
\indent In this macroscopic and out-of-equilibrium model experiment, a labyrinthine phase is obtained by applying a magnetic field to a confined granular gas, by the means of externally controlled dipolar interactions. We describe and quantitatively characterize the transition from a gas-like phase towards a globally disordered solid phase. It appears as a three-dimensional effect occurring in a quasi-two-dimensional system. The parameters setting the confinement, the gap $h/a$ and the area fraction $\phi$, are thus essential to explain the phase diagram of this granular medium, as it was also shown for colloidal and hard spheres monolayers~\cite{SchmidtPRE1997,OstermanPRL07,Dobnikar2008}.
Although not presented here, after a fast increase of $B$, \textit{i.e.} a magnetic quench, the labyrinthine phase exhibits a slow dynamics characterized by a slow evolution of its structural properties~\footnote{The slow dynamics reported in some attractive granular gases~\cite{Blair2003,Oyarte2013} results from a cluster growth process and should not be described as an aging phenomenon.}. This aging phenomenon should thus be compared to the slow dynamics of structural glasses~\cite{BarratBook,HunterRPP2012,Stillinger2013}.
In the case of labyrinthine domain patterns arising in continuous systems, the analogy with glasses was also reported in the analysis of their globally disordered structure~\cite{SeulPhilMagB1992} and when studying their relaxation~\cite{ReimannPRE2002}. Finally, whereas the structure of the phases  obtained in this macroscopic experiment resembles those found at thermal equilibrium in Monte-Carlo simulations and colloidal monolayers~\cite{OstermanPRL07,Dobnikar2008,Messina2015}, kinetics of the transition described here is intrinsically an out-of-equilibrium process, which deserves further studies.

\begin{acknowledgments}
The authors would like to thank Leonardo Gordillo, Jean-Claude Bacri and Nicolas Vandewalle for discussions, and Vincent Dupuis for performing particle magnetization measurements. This work has been supported by Universit\'e Paris Diderot and by ESA Topical Team on granular materials No. 4000103461.
\end{acknowledgments}

\section*{Appendix A: Potential magnetic energy for two parallel dipoles with finite magnetic permeability}
For two ferromagnetic spheres of identical diameter $a$ immersed in an unidirectional vertical magnetic field of intensity $B$ and separated by a distance $r$, the potential energy of magnetic interaction reads in spherical coordinates as the interaction of two vertical magnetic dipoles~\cite{Jackson}:
\begin{equation}
U_m (r,\theta) = -\frac{\pi}{16} \frac{{B}^2}{\mu_0} \frac{a ^6}{r^3} 
\left( \frac{\mu-\mu_0}{\mu + 2\mu_0} \right)^2  \left( 2\cos ^2\theta -\sin ^2\theta \right)
\label{eq:Em2s}
\end{equation}where $\theta$ is the polar angle between the two dipoles%\rr{(see Fig.~\ref{fig:f1s})}, 
, $\mu_0=4\pi10^{-7}$\,H.m$^{-1}$ is the vacuum permeability and $\mu$ is the intrinsic magnetic permeability of the sphere material. The induced magnetic fields of the neighboring particles are assumed negligible in front of the external magnetic field. For free moving particles $\theta$ is taken equal to $\pi/2$, whereas for particles belonging to chains $\theta$ is computed from the measured vertical position of the particles (top or bottom). The magnetic potential energy per particle $E_m$ is computed as an average of $U_m$ over interacting pairs of particles~\cite{MerminodEPL2014} and in the limit of large $\mu$. This approximation holds for soft and linear ferromagnetic materials~\cite{Mehdizadeh2010}, which is the case for our particles. 

\section*{Appendix B: Generation of the vibrated and interacting system of particles}
The particles are chromed alloy steel (AISI 52100) spheres of a diameter $a=\nobreak 1$\,mm and of a mass $m=\nobreak 4.07\times10^{-3}$\,g. Using a vibrating sample magnetometer, the magnetization of one particle was measured by V. Dupuis. The magnetic permeability $\mu$ verifies $122 <\mu /\mu_0 < \infty$, in the linear domain ($ -2000\,\text{G}<\nobreak B<\nobreak +2000\,\text{G}$) and the remnant magnetic field $ B_r$ is below $12$\,G. The coercive field is small compared with the values of magnetic field $B$ used in our experiments. Within this range of $B$, the response to magnetic excitation is linear.
The square aluminum cell (side $9$\,cm long) containing these particles (see Fig.~\ref{fig:f1}) is vertically driven by an electromagnetic shaker. The forcing is sinusoidal at the frequency $f_0=300$\,Hz and the root mean square (RMS) acceleration of vibration is fixed at $\Gamma=\nobreak 21.9\,$m.s$^{-2}=\nobreak 2.23\,g$ for all experiments, with $g$ the gravity acceleration. This value corresponds to the upper limit of the linear response domain of the granular temperature $T_g=\nobreak E_c/m$ as a function of $\Gamma$~\cite{ReisPRE2007}. The two coils are a Helmholtz pair current-controlled supplied generating a nearly homogeneous vertical magnetic field $B$ across the cell (3\% of variation measured). Immersed in this magnetic field, the particles are magnetized into induced dipoles vertically oriented (particle rotating velocity is negligible compared with the speed of the magnetic domains rearrangements). 

\section*{Appendix C: Particle detection}
An annular light-emitting diode (LED) array above the cell produces a high-contrast circular signal on the chromed particles whose positions are recorded from above using a high-speed video camera at high resolution ($1152\times1152$\,pixels at $780$\,Hz). The region of interest $S$ is $5.7$\,cm$\times 5.7$\,cm around the cell center (see Fig.~\ref{fig:f1}). The particle diameter is about $20$\,pixels. For individual particle detection, we used a convolution-based least-square fitting routine~\cite{ReisPRE2007,ShattuckWeb} completed by an intensity-weighted center detection algorithm (accuracy estimated less than $0.3$\,pixel). Particle trajectories  were reconstructed using a tracking algorithm~\cite{Crocker,CrockerWeb}. 

\section*{Appendix D: Experimental protocol}
It is fully automated for the sake of robustness. Every single experiment is non-correlated with the others. The amplifiers of the electromagnetic shaker and of the Helmholtz coils are computer-controlled \textit{via} a data acquisition card. The experimental protocol routine is written in Matlab. It also proceeds to the dialogue with the camera, i.e., configuring and starting the video recordings, as well as to the recording of the data from the accelerometer and the Hall effect sensor.
All experiments are performed according to the following protocol. First the shaking is activated ($\Gamma=21.9\,$m.s$^{-2}$) while the magnetic field remains zero. The magnetic field is then linearly increased (rising rate $\alpha_q\equiv \text{d}B/\text{d}t$ is kept fixed for all experiments) up to its higher plateau value~$B$. Afterwards, a waiting time is respected prior to proceeding to the recordings. It is chosen along with the recording time length so as to reach the chosen mean aging time~$\tau_w$. In all the experiments presented here, $\alpha_q=1\,$G.s$^{-1}$, $\tau_w=30\,$s and recordings last at least $2$\,s. Note that these two parameters $\alpha_q$ and $\tau_w$ have a noticeable influence on the nature of the labyrinthine state reached for high values of $B$, implying that a slow dynamics is at play.\\
\indent For the $5000$ particles introduced in the experiments, the area fraction evaluated on the cell is equal to $0.485$. However, within the region of interest $S$, as the boundaries are not repulsive~\cite{MerminodEPL2014}, $\phi$ decreases from $0.58$ to $0.46$ with $B$ until $B \approx 80\,$G. From the appearance of the first chains, $\phi$ remains nearly constant.

%%%%%%%%%%%%%%%%%%%%%%%%%%%%%%%%%%%%%%
%%%%%%%%%%%% REFERENCES %%%%%%%%%%%%%%%%%%
%%%%%%%%%%%%%%%%%%%%%%%%%%%%%%%%%%%%%%
%

%\bibliography{LG15612E_Merminod_new}

\end{document}